# Slow Dynamics in Glasses

J. C. Phillips,

Dept. of Physics and Astronomy, Rutgers University, Piscataway, N. J., 08854-8019

ABSTRACT

Minimalist theories of complex systems are broadly of two kinds: mean-field and axiomatic. So far all theories of complex properties absent from simple systems and intrinsic to glasses are axiomatic. Stretched Exponential Relaxation (SER) is the prototypical complex temporal property of glasses, discovered by Kohlrausch 150 years ago, and now observed almost universally in microscopically homogeneous, complex non-equilibrium materials, including luminescent electronic (Coulomb) glasses. Critical comparison of alternative axiomatic theories with both numerical simulations and experiments strongly favors dynamical trap models over static percolative or energy landscape models.

## 1. Introduction

Glasses present unique challenges to computation and theory: if microscopically homogeneous, well below the glass transition temperature $T_g$ they are in metastable states that last indefinitely, yet they exhibit no discernible symmetry. They do exhibit well-defined phase transitions, and some of these appear to be generic to many glasses, not only molecular, but also electronic as well [1]. Explicit *off-lattice* Newtonian molecular dynamics simulations (MDS) have proved to be very difficult for all but the simplest "glasses" (mixed Lennard-Jones (LJ) systems [2,3]), for which no physical examples are known to be metastable for laboratory times > a few ns. Glass formers exhibit two ubiquitous properties: diverging viscosity $\eta \sim \exp([T^*/(T - T_g)]^\alpha)$ with $\alpha \sim 1$ (or 1.5 [4]) in supercooled liquids on approaching $T_g$; and stretched exponential relaxation (SER) of a pulse I according to $I(t) \sim \exp[-(t/\tau)^\beta]$ near and below $T_g$. Modern digital techniques, combined with much longer excited lifetimes in glasses than in liquids, have made it possible to determine the dimensionless stretching fraction $0 < \beta(T_g) = \beta_g < 1$ with accuracies $\sim$ 1-2%. Both the phase diagrams and the relaxation dynamics of glasses have been measured with much higher accuracy in glasses than in most supercooled liquids, presenting theory with challenging data.

**2. Exponentially slow dynamics**

Broadly speaking there are three schools of thought concerning the dynamics of supercooled liquids and glasses. (1) Hydrodynamic or mode-coupling theory is a polynomial model which attempts (with mixed success [5]) to extrapolate properties of the liquid near its melting point to predict $T_g$; it cannot explain glassy exponential dynamics. (2) Landscape models [3,6] represent glassy disorder through static models, for example, Gaussian distributions of trap activation energies obtained from molecular dynamics simulations (MDS) on LJ glasses; with further approximations, these give SER with $\beta_g$ errors > 30%. (3) Abstract axiomatic models derived from the classic or standard Scher-Lax (SL) electronic trap model based on dispersive electronic transport. These yield SER exactly in the asymptotic limit of long times, with explicit expressions for $\beta$ which can be compared directly to experiment on a great range of glass formers, and also permit discussion of variations in $\beta$ with probe [7].

Traps enter the derivation of SER differently in landscape and axiomatic models. Landscape models are static: one examines some property of a specified configuration, such as a distribution of fictive activation energies, and somehow extracts SER. Axiomatic models exploit the fact that a simple dynamical model exists for which SER is asymptotically exact. This is the SL model first introduced to describe dispersive (not ballistic) transport in amorphous semiconductors [8,9], and subsequently refined by many authors [7] to describe SER in many molecular and electronic glasses. Relaxation occurs by diffusion of excitations to fixed, randomly distributed sinks; with the passage of time excitations must diffuse longer distances to reach sinks.

Two kinds of landscape models produce SER. If relaxation is defined by an autocorrelation function, then random walks [10] on closed hyperspheres give simple exponential relaxation ($\beta = 1$). One can represent glass relaxation by a random walk on the backbone of a percolative network in a fractal hyperspherical n = (d – 1) dimensional surface or Euclidean volume configuration space of dimension d; this leads to SER with $\beta_P(d)$ decreasing smoothly from (0.70, 0.53, 0.44) for n = 2,3,4 to its mean-field limit 1/3 for d = n -1 $\geq$ 6. So far there appears to be no correspondence between the static values of d or n and experimental $\beta_g$ 's.

The landscape terminology is popular for analyzing slow dynamics of supercooled LJ liquids [2,3], where one finds a distribution of fictive activation energies E between "metabasins". This distribution appears to be approximately Gaussian, which leads to a distribution $\psi_G(\tau)$ of trap times $\tau$,

$$\psi_G(\tau) \sim \exp[-(\ln(\tau/\tau_0))^2/\Delta] \qquad (1)$$

At first $\psi_G(\tau)$ seems very different from the $\psi_S(\tau)$ distribution for SER ($\alpha = \beta/(1 - \beta)$),

$$\psi_S(\tau) \sim \exp\{-C(\beta)\exp[\alpha \ln(\tau/\tau_0)]\} = \exp\{-C(\beta)(\tau/\tau_0)^\alpha\} \qquad (2)$$

which reduces to a simple exponential ($\alpha = 1$) for $\beta = \frac{1}{2}$. It is never a Gaussian in $E \sim \ln\tau$, but it is (somewhat "accidentally") a Gaussian in $\tau$ for $\beta = 2/3$. In any case, comparing (2) with (1), the double exponential here suggests that the "independent" Gaussian excitations in E [3] are even more strongly correlated with each other in the glass than in a liquid.

A series expansion [3,6] of the exponential argument of (1) yields SER approximately with $\beta = \beta_G = (1 + \Delta^2/2)^{-1/2}$. From MDS on LJ glasses one obtains independent estimates [3] of $\Delta$, from which one can calculate $\beta_G$. There have been two MDS on mixed LJ liquids [2,3] that yield estimates of $\beta_g = \beta(T)$ as $T \to T_g$. They agree that $\beta_g(d = 3) = 0.61(1)$, corresponding to $\beta_P(3) = 0.53$ and $\beta_G(3) = 0.41$. Comparing these results, it appears that the static approximation alone, common to the percolative [10] and Gaussian [3] models, underestimates $\beta_g(3)$ by about 12%, while the additional fictive "landscape" approximations involved in forcing (2) from (1) leads to a further 20% reduction [3]. As we shall see below, a change in $\beta$ from 0.6 to 0.4 implies a very large and basic change in relaxation mechanisms. Neither static model suggests a systematic procedure for interpreting experimental data with accuracies of a few %.. Historically, Gaussian models have long been considered unacceptable in fitting even relatively limited dielectric relaxation data [11]; they have not been used recently to fit any temporal relaxation data (numerical or experimental).

The axiomatic model provides a clear path to understanding glassy dynamics; it has correlated and consolidated data up to 1996 for both electronic and molecular glasses, and has successfully predicted much subsequent data. For short-range forces only it predicts $\beta_g(d) = d/(d+2)$, with $\beta_g(d=3) = 3/5 = 0.60$, in excellent agreement with MDS on mixed LJ liquids (above). The result $\beta_g(d) = d/(d+2)$ follows directly from dimensional analysis of the diffusion equation. The diffusion equation does not separate motion into static activated and dynamic ballistic parts, and thus it automatically includes medium dynamics that are associated with backflow and recoil in conventional kinetic calculations.

## 3. Comparison with experiment: molecular glasses

The diffusive sink model successfully [12] describes (within 1-2%) the short-range part of the modern molecular glass relaxation data base for $\beta_g = 3/5$ for 9 inorganic network glasses, 6 nonpolar organic glasses and 5 alcohols [7]. This success is apparently contradicted by its striking failures to explain the values of $\beta_g$ for many other glasses, especially a-Se and 8 fully noncrystalline polymers. There one finds that the values of $\beta_g$ are all clustered around $\beta_g = 0.43$. (Some well-known commercial polymers, like polyvinyl chloride (PVC) and polystyrene (PS) have much lower values of $\beta$, but it has been known since the 1930's that PVC and PS are partially crystallized and microscopically inhomogeneous (atactic) [7].) However, let us insert 0.43 in $\beta_g(d^*) = d^*/(d^*+2)$, and solve for $d^*$. The result is $d^* = 3/2$, and at this point one would have to be deaf not to hear what experiment is trying to tell us. While $d = 3$ suggests relaxation by density fluctuations, there are long-range uniaxial strain fields associated with the chain structures of polymers and a-Se. One can then suppose that the glassy relaxation dynamics is affected both by short-range forces, which are effective, as well as long-range strain forces, which are ineffective, and define an effective or fractal dimensionality $d^*$ for competitive relaxation involving short- and long-range forces, which is $d^* = fd$, where $f = ½ =$ (number of short-range forces)/(total number of forces) and

$$\beta_g(d^*) = d^*/(d^* + 2) = d/(d + 2/f). \quad (3)$$

Numerical studies have shown [7] that SER is an asymptotic property that sets in only after a transient simple exponential has died out and many-body correlations dominate. If one starts out with a mathematical model of sinks randomly embedded in an ideal (random) gas, then this asymptotic regime is reached only after astronomically long times, when the signal is exponentially small ($\sim 10^{-10}$ or less). However, simulations of good glass-formers typically reach the asymptotic regime when the signal is ~ 3/4 of the initial value [7]. The key to understanding this paradoxical result is that to avoid crystallization and reach a glass transition, one must *choose* a good glass-former to study, and such a good glass former retains only a small fraction (such as 1/4) of the truly "random" disorder contained in the mathematical gas model; the chosen model by selection (usually based on extensive numerical surveys) already contains a high level of configurational entanglement, designed to simulate a good glass-former.

The differences between the predictions of Eqn. (3) and the random walk percolative model [10] are instructive. Both predict that the combination of short-range and long-range forces reduces β (wider distribution of relaxation times); a plausible value for d in the percolative model would be d = 6, which is in the mean-field limit with $\beta_P = 1/3$. When the ineffectiveness of long-range forces in relaxation by density fluctuations is noted, (3) gives $\beta_g = 3/7$, in excellent agreement with experiment, and significantly different from 1/3..

Recent molecular data dramatically confirm the predictions of the axiomatic configurational sink model for $\beta_g(d^*)$ [13]. The most-studied non-polar organic is OTP (o-terphenyl), which is available commercially in high purity; as expected, its relaxation is dominated by density fluctuations [7] and $\beta_g = 0.61(1)$. There have been two recent studies of OTP, one by Brillouin scattering [14], which excites surface waves that relax by a mixture of short- and long-range forces, and these gave $\beta_g = 0.43$, as predicted. A new method, multi-dimensional NMR, enables one to study [15] the relaxation of states prepared over long and short times, so that relaxation can take place either with short-range forces only, or with a mixture of short-and long-range forces, leading to two values of $\beta_g$, 0.59 and 0.42, both in excellent agreement with the predicted values of 3/5 and 3/7.

Photon correlation spectroscopy revealed [16] the differences between relaxation of poly(propylene glycol) (PPG) and PG in the wide time range $10^{-8}$–$10^{-4}$ s. In PPG hydroxyl terminals are separated by N organic nonpolar monomers, and for 7 < N < 70 these polymers exhibit a constant β nearly independent of N with β = 0.43, in good agreement with the value 0.42 obtained by stress measurements [2], and quite distinct from the N= 1 PG value [2] $β_g$ = 0.61. Thus the value of $β_g$ predicted for mixed short- and long-range forces of 3/7 is confirmed for PPG, while PG itself exhibits relaxation by short-range forces only, with $β_g$ = 3/5, with a smooth cross-over between the two limits.

**4. Comparison with experiment: luminescent electronic glasses**

There is nothing about the axiomatic model that is specific to molecular glasses, so it is not surprising that its most spectacular successes occur when the values certified by molecular examples are tested against electronic glasses. Figs. 1(a)-(d) show data for four unconventional glasses: (a) when band-edge carrier relaxation is measured [17] by studying charge sweep-out in a-Si:H stored metastably at room T, the lowest value of β (identified as $β_g$) is about 0.44 at T = 300K; the observed T dependence extrapolated linearly to higher T fitted also H ion diffusion; this common behavior provides strong support for the trap model not only for electronic relaxation, but also for molecular relaxation as well. (b) When optically excited band edge carriers relax in $C_{60}$, β = $β_g$ = 0.40(5) is independent of T, because all the states in the excited band are localized at energies 4 eV below the mobility edge [18]. (c,d) A much richer T dependence of photoinduced absorption occurs in commercial filters composed of Cd(S,Se) nanocrystallites embedded in a borosilicate matrix [19], in both the virgin (d) and annealed (c) samples. The nanocrystallites are enlarged and faceted by annealing, and the oscillations at low T are related [7,19] to these changes. The carriers are apparently mobile at high T, with a glass transition near T = 300K (probably designedly, as this minimizes spectral thermal and strain broadening at the normal operating temperature), again with β = $β_g$ = 0.40(5). In all these examples it appears that at the glass transition relaxation occurs through a combination of short-range tunneling and long-range

Coulomb interactions, in other words, the systems form Coulomb glasses with $f = ½$ in eqn. (3).

SER is a powerful tool for studying the metal-insulator transition (MIT) and the Coulomb glasses formed at this transition, as shown by recent femtosecond pump-probe experiments on porous Si [20]. Porous Si, anodized to produce a thin oxide, is a complex material; at first it appears unsuitable for a theory that requires microscopic homogeneity. However, at the MIT carriers are distributed over a percolative backbone, and one can ask whether the dynamical properties of the carriers are dominated by either the static nature of the percolative geometry [10], or simply by dynamical competition between effective and ineffective relaxation channels, as in Eqn. (3). Recovery of photodarkened films exhibits SER with $\beta(\omega)$ reaching a minimum, Fig. 2, at the optical MIT energy of 2.65 eV. The minimum value of $\beta$, obtained by fitting the recovery data, also illustrated in Fig. 2, with a transient simple exponential and a residual stretched one, is $\beta_g = 0.40(3)$, in very good agreement with the Coulomb glasses of Fig.1 and the predicted (short, long) competitive value of $\beta = 3/7$. These digital data are much more accurate than one might suppose from only glancing at Fig. 2; there much of the "scatter" is due to modulation by coherent phonons, whose spectra are deconvoluted [20] with a resolution of ~ 10 cm$^{-1}$. SER of luminescence has also been reported for porous $Si_{1-x}Ge_x$ alloys; for x = 0.05 and 0.30 the smallest values of $\beta$ are observed near luminescence energies of 2.1(1) eV. Again these are $\beta = 0.40(5)$, very similar to porous Si [21], although the radiative defects involved may be different. Taken altogether, these data suggest that the host region of the luminescent defects is topologically the same as the region that dominates the optical MIT.

The most natural candidate for the optical MIT percolative backbone is an SiO surface monolayer (or an SiO monolayer interface with a thin $SiO_2$ oxide layer), which is microscopically homogeneous because it is nearly strain-free [1]. This topological mechanism provides a more sophisticated explanation of the anomalous, highly material-specific luminescence of porous Si than any of the alternative *non-specific* geometrical mechanisms [20]. The growth of a thicker oxide stresses the SiO interface and creates additional nonradiative interfacial defects that reduce the luminescence efficiency and shorten the luminescence lifetime by several orders of magnitude [21]. There is no evidence for enhanced luminescence in $GeO_x$ [22], which agrees with theory [7], which says that neither g-$GeO_2$ nor the Ge/$GeO_2$ interface are microscopically homogeneous enough to host radiative defects.

Not all electronic systems exhibit glassy properties associated with Coulomb frustration. Specifcally, photoluminescence in sufficiently homogeneous Si nanodots with carriers confined by $SiO_2$ (but not SiH) surfaces to dimensions small compared to 5 nm [23] should exhibit SER with β close to 3/5 (short range forces only). Indeed, SER has been observed [24] in ~ 4 nm Si nanodots embedded in $SiO_2$, with β = 0.57(3), in excellent agreement with the predictions of the SLP model.

Of course, many electronic experiments on putative Coulomb glasses do not exhibit SER because the samples are not sufficiently homogeneous in the given context. Similarly, so far the very formidable computational problems associated with long-range Coulomb interactions have prevented even Monte Carlo random lattice simulations from going far enough beyond the transient temporal region to cover the residual region where SER dominates and is identifiable [25]. Agreement between extrinsic non-exponential transients obtained theoretically and equally extrinsic microscopically inhomogeneous non-exponential behavior observed experimentally is usually not possible. Thus the SER data discussed above contain exceptionally valuable evidence for collective behavior in Coulomb glasses; so far only the axiomatic model has been able to describe this collective behavior and predict $β_g$ = 0.4.

**5. Supercooled liquids**

An intriguing question is whether one should first study supercooled liquids or melts in order to understand glasses and glass formation, or whether the reverse procedure is more appropriate. Here we have found that the chemical trends of SER in glasses practically explain themselves in the context of the SLP diffusion to sinks model. As T decreases towards $T_g$ in the supercooled liquid, β(T) decreases towards $β_g$ = β($T_g$), and it is β($T_g$) that exhibits *universal* values. The relaxation channels that occur in the SLP model are similar to the constraints that have proved so effective in understanding the phase diagrams and the origin of the intermediate phase in both electronic and molecular glasses [1]. The patterns of β(T) in the supercooled liquid are *variable*: sometimes there is a plateau before T reaches $T_g$, but often there is not. The edges of the intermediate phase are *abrupt* in both electronic and molecular glasses, but they are *thermally broadened* in supercooled liquids [1], so much so, that the intermediate phase *was not identified* in the earlier data on supercooled liquids.

Overall, metastable glassy phases behave much like low temperature crystalline phases; these are generally understood more accurately than the higher temperature crystalline phases. Glass-forming liquids probably are still exponentially complex, which explains the mixed success of polynomial mode-coupling theory [5] in predicting $T_g$, and its inability to explain SER in supercooled liquids. However, in the liquid, glassy constraints are partially broken, presumably hierarchically (as in proteins [1]). In favorable cases where constraints can be connected directly to spectroscopic vibrational bands, one might be able to predict β(T) in the supercooled liquid, but so far other problems have seemed to be more important and more accessible.

## 6. Conclusions

The axiomatic diffusion to sinks model (with effective and ineffective diffusion channels) is somewhat abstract, but it exhibits SER without the kinds of entanglements (especially with long-range forces) that arise in network structures and have made the residual time range inaccessible to MDS. Apart from mixed LJ liquids and spin glasses [2,3,7], SER has also been obtained in several abstract lattice models [26,27] where dynamic clustering effects appear, similar to those that are well known in network glasses [12,28], and which are now being studied by MDS [29]. The overall consistency of the data for $β_g$ with the predictions of the trap model leaves little doubt that this is the correct model for describing SER in microscopically homogeneous materials. Its success in the context of luminescent electronic glasses is especially significant, because the ms time scales of luminescence are similar to the ms time scales of protein dynamics.

The static landscape model is widely invoked in qualitative descriptions of protein dynamics, but the present analysis shows that it is unsuitable for quantitative analysis of slow dynamics (such as protein folding). In the context of biomaterials, it is likely that intrinsic SER has been observed [30] in trehalose, with β = 0.38(2). Ordinarily it is difficult to achieve microscopic homogeneity in biomaterials, but trehalose has unique properties: it is a nonreducing glycoside that has been found in large quantities in organisms (algae, bacteria, fungi, insects, invertebrates, and yeasts as well as a few flowering plants) that are able to survive extreme external stresses such as high or very low temperatures or periods of complete drought. These qualities led to the suggestion [31], now apparently prescient, that trehalose forms a glassy structure around embedded biomolecules and inhibits thus the denaturation due to formation of ice crystals. One can also examine the fast dynamics of trehalose [32], which reflects directly the stability of the hydrogen bond network, either by neutron scattering or by MDS, compared to the slow dynamics (15 decades slower!) measured by viscosity. Here we can see that relaxation of trehalose is suppressed at short times relative to long times in much the same way as with other molecular systems with mixtures of short-and long-range forces (such as Se, for which the values of β are indeed observed to be the same, even when separated by 15 decades [7]). In fact, packaging by trehalose [(4 H bonds)/(6 covalently bonded atoms)] is optimized when it is diluted with 5% glycerol [(3 H bonds)/(3 covalently bonded atoms)], which alters the balance of forces slightly in favor of H bonds; this dilution presumably also suppresses nanocrystallization. The special packaging properties of trehalose show similarities to those of Si/SiO$_2$ interfaces, and can be treated by similar topological methods; they deserve further discussion elsewhere.

**Figure Captions**

Fig. 1. (a) Stretching fraction $\beta(T)$ of carrier relaxation after charge sweep-out in a-Si:H (low T filled circles) and H ion diffusion (higher T open circles and filled square) [17]. The minimum value of $\beta$ at the sample storage (room) temperature (fully relaxed H configuration) is 0.45(5). The linear T line shows that the dynamics of relaxations of n-type carriers to trapped holes and H ions to traps (possibly voids) share common functionality. (b) In $C_{60}$, band edge carrier $\beta = \beta_g = 0.40(5)$ is independent of T (see text) [18]. (c) and (d): $\beta(T)$ of Cd(S,Se) nanocrystallites embedded in a borosilicate matrix [19], in both the virgin (d) and annealed (c) samples. Apart from oscillations associated with nanocrystallite morphology, $\beta = \beta_g = 0.40(5)$.

Fig. 2. Upper panel: stretching fraction $\beta(T)$ of carrier relaxation in porous Si following femtosecond laser pulse at 2.34 eV. As in Fig. 1, the minimum value is $\beta_g = 0.40(3)$, characteristic of a Coulomb glass. Lower panels: Transmission and reflectivity changes after femtosecond laser pulse at 2.34 eV, probed at 2.50 eV. Relaxation data at all energies are modulated by coherent phonons [20].

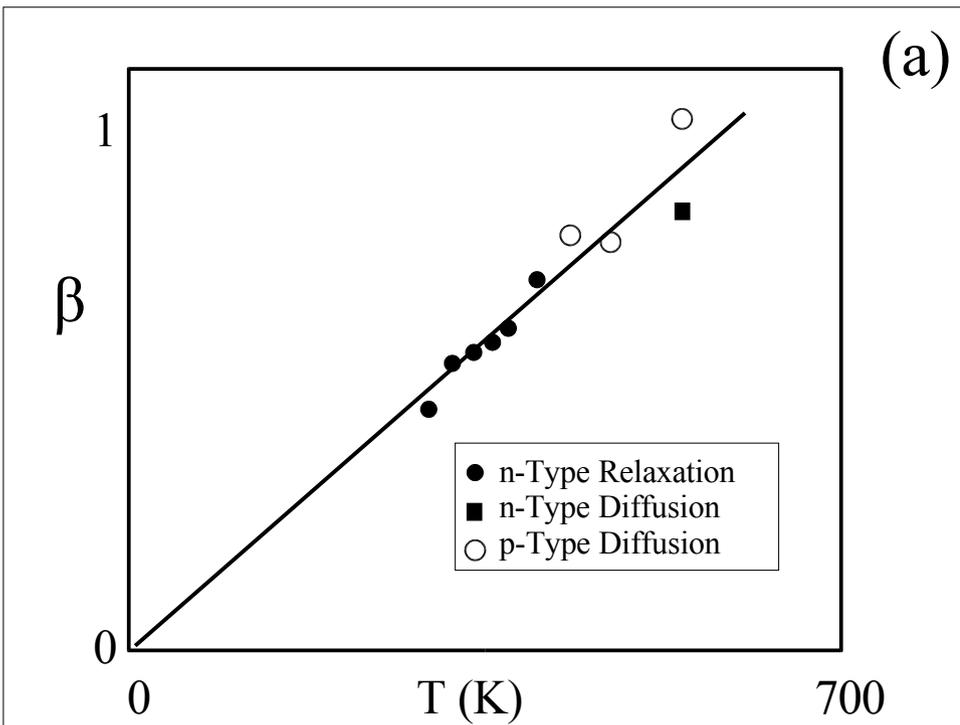
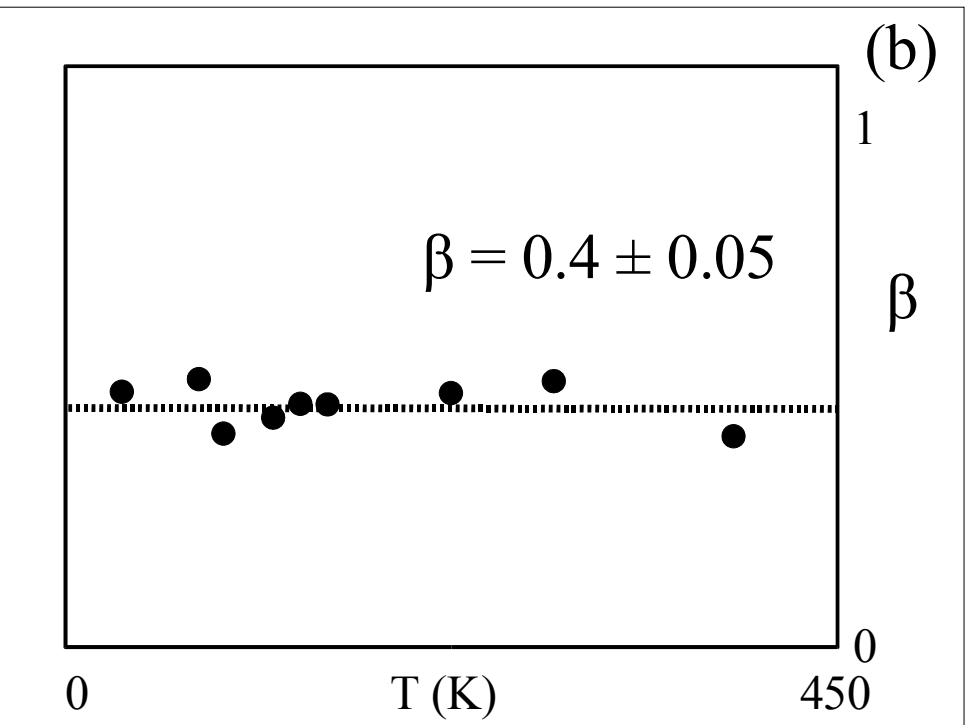
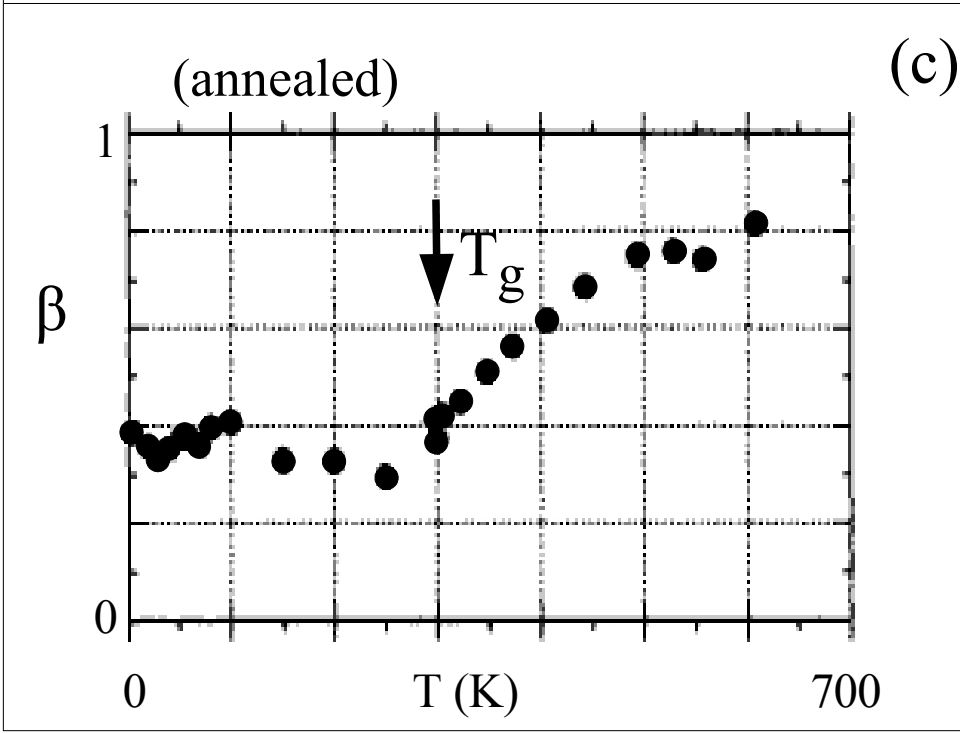
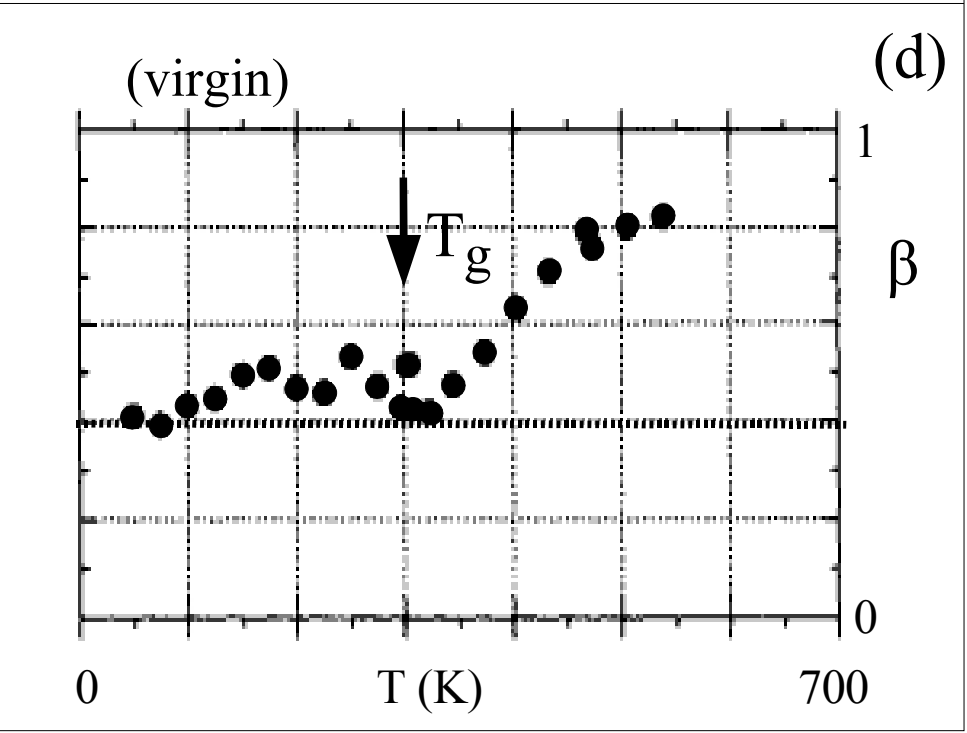

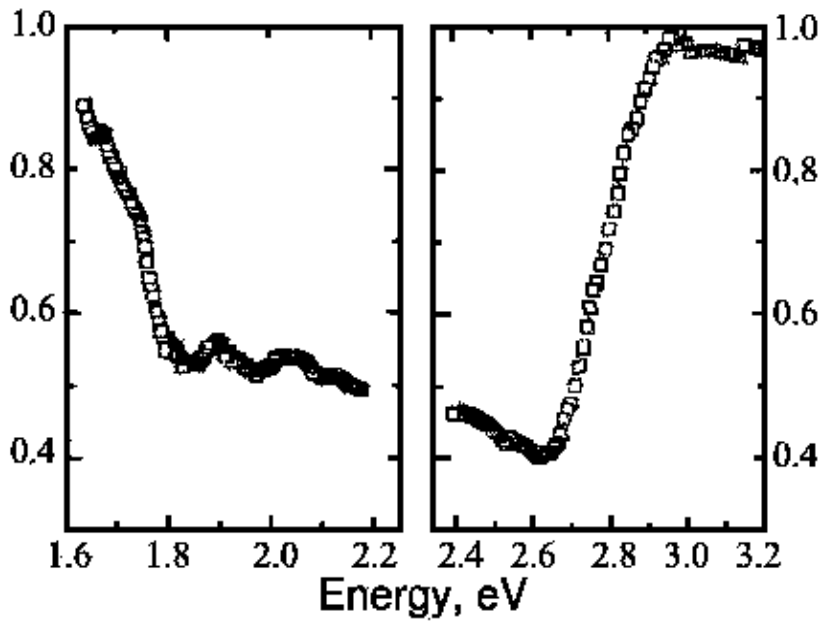

β(ω)

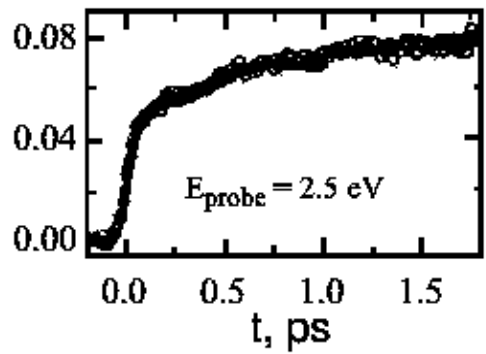

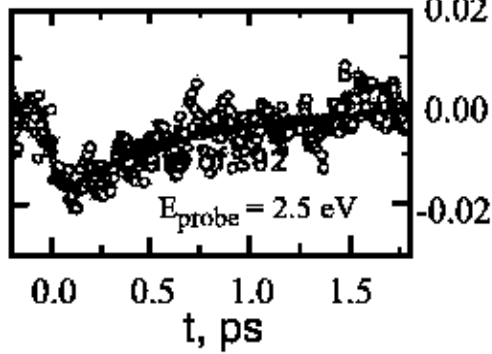